\documentclass[aps,twocolumn,superscriptaddress,raggedbottom,showpacs,floatfix]{revtex4-1}
\usepackage[pdftex]{graphicx}
\usepackage{amssymb}
\usepackage{amsmath}
\usepackage{color}
\usepackage{hyperref}
\hypersetup{colorlinks=true,linkcolor=blue,urlcolor=blue,citecolor=blue}

\begin{document}

\title{Dielectric response and novel electromagnetic modes\\ in three-dimensional Dirac semimetal films}
\author{O. V. Kotov}%
\affiliation{Institute for Spectroscopy, Russian Academy of Sciences, 142190 Troitsk, Moscow, Russia}%
\affiliation{Dukhov Research Institute of Automatics (VNIIA), 127055 Moscow, Russia}%
\author{Yu.E. Lozovik}%
\email[Corresponding author: ]{lozovik@isan.troitsk.ru}%
\affiliation{Institute for Spectroscopy, Russian Academy of Sciences, 142190 Troitsk, Moscow, Russia}%
\affiliation{National Research University Higher School of Economics, 101000 Moscow, Russia}
\affiliation{Dukhov Research Institute of Automatics (VNIIA), 127055 Moscow, Russia}%
\begin{abstract}
Using the Kubo formalism we have calculated the local dynamic conductivity of a bulk, i.e., three-dimensional (3D), Dirac semimetal (BDS). We obtain that at frequencies lower than Fermi energy the metallic response in a BDS film manifests in the existence of surface-plasmon polaritons, but at higher frequencies the dielectric response is dominated and it occurs that a BDS film behaves as a dielectric waveguide. At this dielectric regime we predict the existence inside a BDS film of novel electromagnetic modes, a 3D analog of the transverse electric waves in graphene. We also find that the dielectric response manifests as the wide-angle passband in the mid-infrared (IR) transmission spectrum of light incident on a BDS film, which can be used for the interferenceless omnidirectional mid-IR filtering. The tuning of the Fermi level of the system allows us to switch between the metallic and the dielectric regimes and to change the frequency range of the predicted modes. This makes BDSs promising materials for photonics and plasmonics. 
\end{abstract}

\pacs{71.45.Gm, 73.20.Mf, 78.40.Kc}

\maketitle

\section{Introduction} \label{Sec1}

A great attention has recently been attracted to Dirac fermion systems by the discovery of graphene and topological insulators (TIs). Graphene is known for its unique electronic and optical properties caused by two-dimensional (2D) Dirac fermions in its electronic structure \cite{Novoselov,CastroNeto}. The main feature of strong three-dimensional (3D) TIs is the coexistence of the bulk energy gap and the topologically protected gapless surface states formed by an odd number of the 2D Dirac fermions with the helical spin texture \cite{Hasan,QiZhang}. Furthermore, opening the gap in the surface states by a time reversal or a gauge symmetry breaking causes a remarkable magnetoelectric effect \cite{QiHughes,Essin}. Recently, the accent in the Dirac systems research shifted to the investigation of a novel state of quantum matter that can be considered as ``3D graphene'' --- 3D Dirac semimetals, also called bulk Dirac semimetals (BDSs). The 3D Dirac nature of the quasiparticles was experimentally confirmed by the angle-resolved photoemission spectroscopy investigation of $\textrm{Na}_3\textrm{Bi}$ \cite{Liu:Science}, $\textrm{Cd}_3\textrm{As}_2$ \cite{Borisenko,Neupane,Liu:Nature}, and $\textrm{ZrTe}_5$ \cite{QLi} and the optical conductivity measurements of $\textrm{Cd}_3\textrm{As}_2$ \cite{Neubauer_opt_exp}, $\textrm{ZrTe}_5$ \cite{Chen_opt_exp}, $\textrm{AlCuFe}$, and similar quasicrystals \cite{Basov_opt_exp}. Though 3D Dirac states in BDSs are not topologically protected as 2D Dirac states on the surface of a TI, they still have crystalline symmetry protection against gap formation \cite{Wan,Fang,Young}. This protection in some samples results in ultrahigh mobility up to $9\times10^6\mathrm{cm^2V^{-1}s^{-1}}$ at 5K \cite{Liang_mobility}, which is much higher than in the best graphene ($2\times 10^5\mathrm{cm^2V^{-1}s^{-1}}$ at 5K) \cite{Bolotin}. Furthermore, theory predicts that each doubly degenerate 3D Dirac point can split into two topologically protected Weyl nodes that are separated in momentum (if time-reversal symmetry is broken) or energy (if space inversion symmetry is broken) spaces, thus realizing a topological Weyl semimetal (WS) phase \cite{Murakami,Wang,Halasz-Balents,Burkov-Balents}. The families of magnetic materials including pyrochlore iridates $\textrm{Y}_2\textrm{IrO}_7$ and $\textrm{Eu}_2\textrm{IrO}_7$ \cite{Wan,Sushkov_opt_exp}, and ferromagnetic spinels $\textrm{HgCr}_2\textrm{Se}_4$ \cite{Xu_spinels}, and nonmagnetic materials including TaAs, TaP, NbAs, and NbP \cite{Huang_arcs,Weng_PRX,Shekhar,Lv_TaAs,Behrends,Xu_TaAs,Xu_NbAs,Xu_TaP,Xu_opt_exp} have been recently predicted and experimentally realized to be natural WSs (the detailed WS classification can be found in the reviews \cite{RevDirac2014,RevDirac2016}). Moreover, exotic quadratic double Weyl fermions and unusual equilibrium dissipationless current induced by an external magnetic field were predicted in $\textrm{SrSi}_2$ \cite{Huang_quadratic}. Nontrivial topology of WSs manifests in the unusual surface states with Fermi arcs \cite{Lee,Sun,Ojanen,Hosur,Potter} and in the chiral anomaly \cite{Adler,Bell-Jackiw,Nielsen-Ninomiya,Aji}, which gives rise to a number of novel physical effects: negative magnetoresistance \cite{HosurQi:rev,Huang_Anomaly,Zhang_anomaly}, anomalous Hall effect \cite{HosurQi:rev,Burkov_Hall}, and chiral magnetic effect   \cite{QLi,HosurQi:rev,MaPesin,Baireuther}. The chiral anomaly also influences an electromagnetic (EM) response \cite{Ashby-Carbotte,HosurQi:optics} and plasmons in WSs \cite{Liu-CX,Burkov-pl,ZhouChangXiao,LvZhang,DasSarma_Tpl1,Kharzeev,DasSarma-Tpl2,Zyuzin,Pellegrino,Rosenstein,Ferreiros,DasSarma_SPP}. The manifestations of the chiral anomaly in a density response of WSs in a magnetic field were studied in Refs.~\cite{Liu-CX,Burkov-pl} and in parallel electric and magnetic fields in Ref.~\cite{ZhouChangXiao}. In Ref.~\cite{LvZhang} the BDS polarization function, the Friedel oscillations specific for BDSs and the BDS plasmon spectrum were calculated. The linear temperature dependent scaling behavior of the BDS conductivity \cite{DasSarma_Tpl1} manifesting in the plasmon dispersion for the both undoped and doped cases was studied in Refs.~\cite{Kharzeev,DasSarma-Tpl2} and observed in Refs.~\cite{Sushkov_opt_exp,Chen_opt_exp}. The existence of the chiral EM waves propagating at the vicinity of the magnetic domain wall in WSs was predicted in Ref.~\cite{Zyuzin}. The existence of helicons in WSs (transverse EM waves propagating in 3D electron systems in a static magnetic field) was predicted in Ref.~\cite{Pellegrino}. Also, the existence of the unusual EM modes with a linear dispersion in a neutral (the Fermi level lies at the Weyl nodes) WS was recently predicted within nonlocal response calculations  \cite{Rosenstein,Ferreiros}. In Ref.~\cite{Ferreiros} it is explained that at low frequencies they propagate with the same velocity as electrons, while at high frequencies they have velocity similar to the speed of light in the material. Recently, the observable signatures of the chiral anomaly in WSs have been predicted in the behavior of the surface-plasmon polaritons (SPPs) \cite{DasSarma_SPP}, the dispersion of which turned out to be similar to magnetoplasmons in ordinary metals.

Here we study the behavior of SPP and EM waves in BDSs (not the WS case) films with the Fermi level higher than the Dirac point and the role of the dielectric response in BDSs. SPPs (see, e.g., Refs.~\cite{Raether,Lozovik,Maier,Bozhevolnyi,Pitarke:RevSPP,Abajo:RevSPP}) are coupled EM and charge density waves which can propagate along a metal or semiconductor surface. Using the Kubo formalism in the random-phase approximation (RPA) we have calculated the BDS local dynamic conductivity and the dielectric function, which being substituted in the solution of the electrodynamics equations for a finite thickness layer yields the dispersion laws of SPP and EM waves in BDS films. As a BDS is a 3D counterpart of graphene one can expect that BDS films can support a 3D analog of the unusual evanescent EM waves in graphene. Due to the gapless electron energy spectrum, in BDSs the contribution of the interband electronic transitions in the dynamic conductivity is significantly enhanced, which in some frequency range causes the imaginary part of the conductivity to become negative and the dielectric function to exceed unity (the dielectric response). In graphene or similar 2D Dirac systems the analogous effect leads to an additional type of surface EM waves, the transverse electric (TE) waves \cite{Mikhailov,Jablan,Stauber}. These waves are weakly bound to the surface but exhibit very low propagation loss \cite{Mikhailov} and an extreme sensibility to the optical contrast between dielectrics sandwiching the graphene layer \cite{Kotov}. We obtain that in BDS films this effect leads to the existence of the waveguide (WG) EM modes inside the sample. Moreover, BDS films combine metal and dielectric properties: at frequencies lower than Fermi energy a metallic response in BDS manifests in the existence of SPP, but at higher frequencies a dielectric response becomes dominated and BDS behaves as a dielectric WG. Notice that the frequency window where EM waves in BDSs or WSs are allowed to propagate was mentioned in Ref.~\cite{Zyuzin}. However, to the best of our knowledge the detailed calculations of possible EM solutions in BDS films have not been made yet. We also calculated optical spectra of light incident on a BDS film. We obtain that the dielectric response manifests as the wide-angle passband in the mid-infrared (mid-IR) transmission spectrum of a BDS film.

\begin{figure}[t]
	\centering
	\includegraphics[width=0.8\columnwidth]{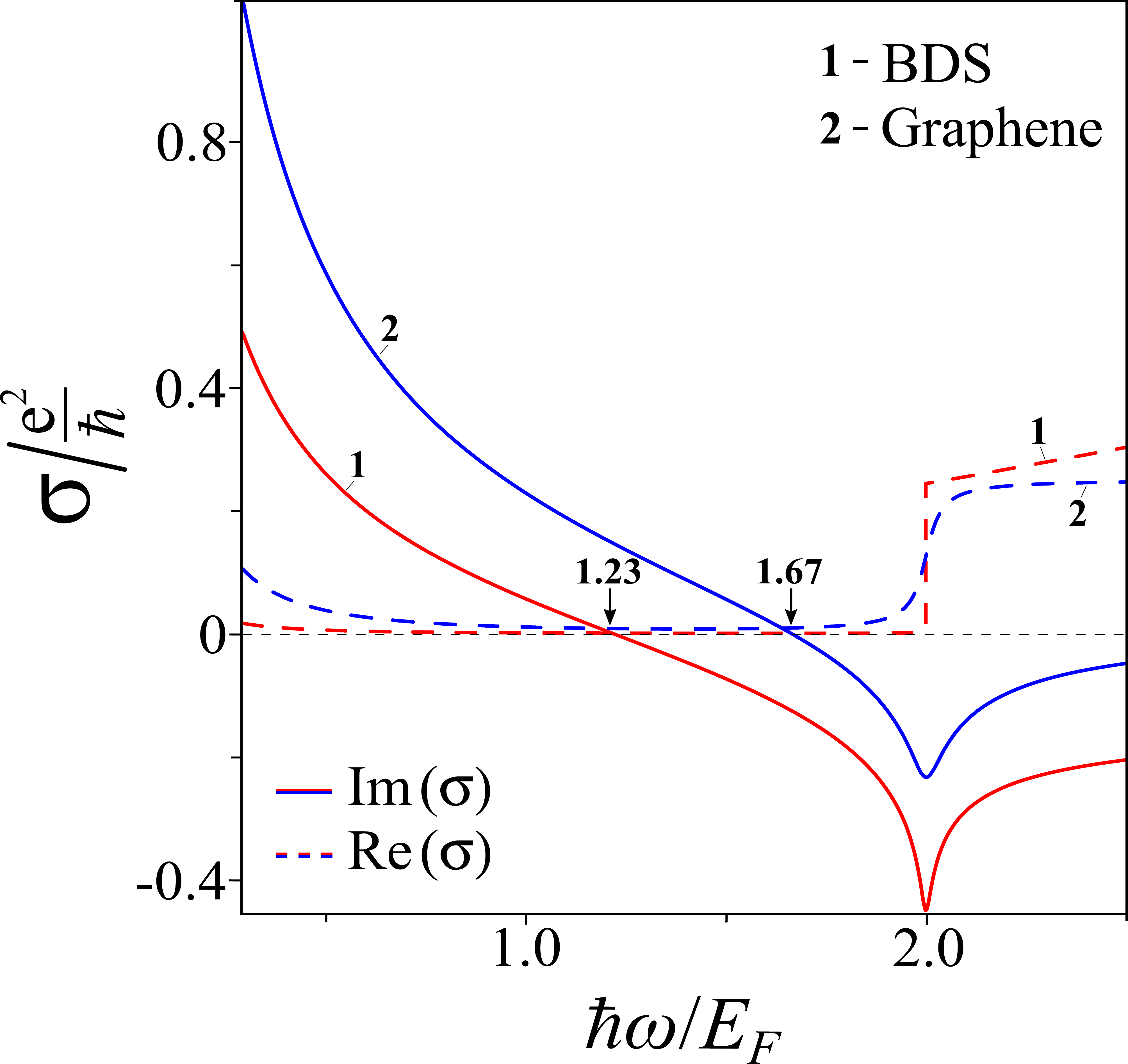}
	\caption{\label{Fig1}(Color online). The real (dash) and the imaginary (solid) parts of the dynamic conductivity for BDS (red (1)) (normalized to 1nm of the thickness) and for graphene (blue (2)) at zero temperature in units $e^2/\hbar$ as a function of the normalized frequency $\hbar\omega/E_F$. The parameters of BDS and graphene are set as $E_F =E_F^G =0.15$eV, $g=40$, $\varepsilon_c=3$, $v_F =v_F^G=10^6$m/s, $\mu =3\times 10^4\mathrm{cm^2V^{-1}s^{-1}}$ ($\tau =4.5\times 10^{-13}$s), $g_G =4$, $\mu_G =10^4\mathrm{cm^2V^{-1}s^{-1}}$ ($\tau =1.5\times 10^{-13}$s).}
\end{figure}

\section{BDS local dynamic conductivity} \label{Sec2}

Using the Kubo formalism in RPA we have calculated at the long-wavelength limit $q\ll k_F$ (the local response approximation) the longitudinal dynamic conductivity of the Dirac 3D electron gas (3DEG) in BDSs. In this work we will not consider the case when BDSs become WSs with the non-zero transverse conductivity and, hence, we will operate only with the longitudinal one. In the case of electron-hole (e-h) symmetry of the Dirac spectrum for the nonzero temperature $T$ we obtain (see Appendix \ref{Appendix_A}):
\begin{align}
\textup{Re}\, \sigma\! \left(\Omega\right )=&\,\frac{e^2}{\hbar}\frac{gk_F}{24\pi}\Omega G\left(\Omega/2\right),\label{ResigT}
\\ \nonumber
\textup{Im}\, \sigma\! \left(\Omega\right )=&\,\frac{e^2}{\hbar}\frac{gk_F}{24\pi^2}\biggl[\frac{4}{\Omega}\left ( 1+\frac{\pi^2}{3}\left(\frac{T}{E_F}\right)^{2}\right)
\\
&\,+ 8\Omega\int_{0}^{\varepsilon _c}\left (\frac{G\left ( \varepsilon \right )-G\left ( \Omega/2 \right )}{\Omega^2-4\varepsilon ^2}\right )\varepsilon d\varepsilon\biggr], \label{ImsigT}
\end{align}	
where $G(E)=n(-E)-n(E)$ with $n(E)$  being the Fermi distribution function, $E_F$ is the Fermi level, $k_F =E_F/\hbar v_F$ is the Fermi momentum, $v_F$ is the Fermi velocity, $\varepsilon =E/E_F$, $\Omega =\hbar \omega/E_F$, $\varepsilon_c =E_c/E_F$ ($E_c$ is the cutoff energy beyond which the Dirac spectrum is no longer linear), and $g$ is the degeneracy factor. At the low-temperature limit $T\ll E_F$ we get:
\begin{align}
\textup{Re}\, \sigma\! \left(\Omega\right)=&\,\frac{e^2}{\hbar }\frac{gk_F}{24\pi}\Omega\,\rm{\theta}(\Omega-2),\label{Resig}
\\
\textup{Im}\, \sigma\! \left(\Omega\right)=&\,\frac{e^2}{\hbar}\frac{gk_F}{24\pi^2}\left[\frac{4}{\Omega }-\Omega\ln\left(\frac{4\varepsilon_c^2}{\left|\Omega^2-4\right|}\right)\right].\label{Imsig}
\end{align}	
Our result for the BDS dynamic conductivity coincides with the expressions for the polarization function $P\left(q,\omega \right)$ calculated in RPA \cite{ZhouChangXiao,LvZhang} at $q\ll k_F$, where $\sigma(\omega )=\frac{ie^2\omega}{q^2}P\left(q\to0,\omega\right)$. For further calculations we will take into account the Drude damping in Eqs.~(\ref{ResigT})--(\ref{Imsig}) by using the substitution $\Omega\to\Omega+i\hbar\tau^{-1}/E_F$, where $\hbar\tau^{-1} =v_F/\left(k_F\mu\right)$ is the scattering rate determined by the carrier mobility $\mu$. The first term in Eq.~(\ref{Imsig}) arises from the intraband conductivity and has the Drude-like form, while the second logarithmic term as in graphene \cite{Mikhailov,Falkovsky} is the negative contribution of the interband transitions (the dielectric response). The real part of the BDS conductivity (\ref{Resig}) also arises from the interband transitions and is responsible for the optical absorption. Unlike graphene, where the absorption is constant, BDSs have the absorption with the linear frequency behavior, as was observed experimentally \cite{Basov_opt_exp,Sushkov_opt_exp,Chen_opt_exp,Neubauer_opt_exp,Xu_opt_exp}. The imaginary part of the BDS conductivity (\ref{Imsig}) differs from the graphene one by the cutoff energy dependence of the logarithmic term and the frequency factor before it. However, as in graphene, in BDSs there is a frequency range where the dielectric response is dominated. Using Eq.~(\ref{Imsig}) we obtain that the imaginary part of the BDS conductivity becomes negative at $\Omega>\Omega_0=1.23$ for $\varepsilon _c=3$ \cite{Ec}, while for the monolayer graphene it becomes negative at $\Omega>1.667$ \cite{Mikhailov}. For the convenience of comparison in Fig.~\ref{Fig1} we plot together the 2D graphene conductivity and the 3D BDS conductivity normalized to 1nm of the thickness. There the dashed lines show the real parts of BDS and graphene conductivities, which at $\Omega>2$ have the linear and the constant frequency behaviors, respectively. Notice that for BDSs this frequency range depends on the cutoff energy (e.g., for $\varepsilon_c =10$ we obtain $\Omega_0=0.91$). Moreover, in BDSs the dielectric response may manifest at frequencies below $\Omega_0$ due to the interband contributions from bands lower than the valence one which are not included in Eqs.~(\ref{Resig}) and (\ref{Imsig}). Thus to analyze the dielectric response more adequately one should consider the dielectric function of BDSs in detail.       

\begin{figure*}[t]
	\centering
	\includegraphics[scale=0.15]{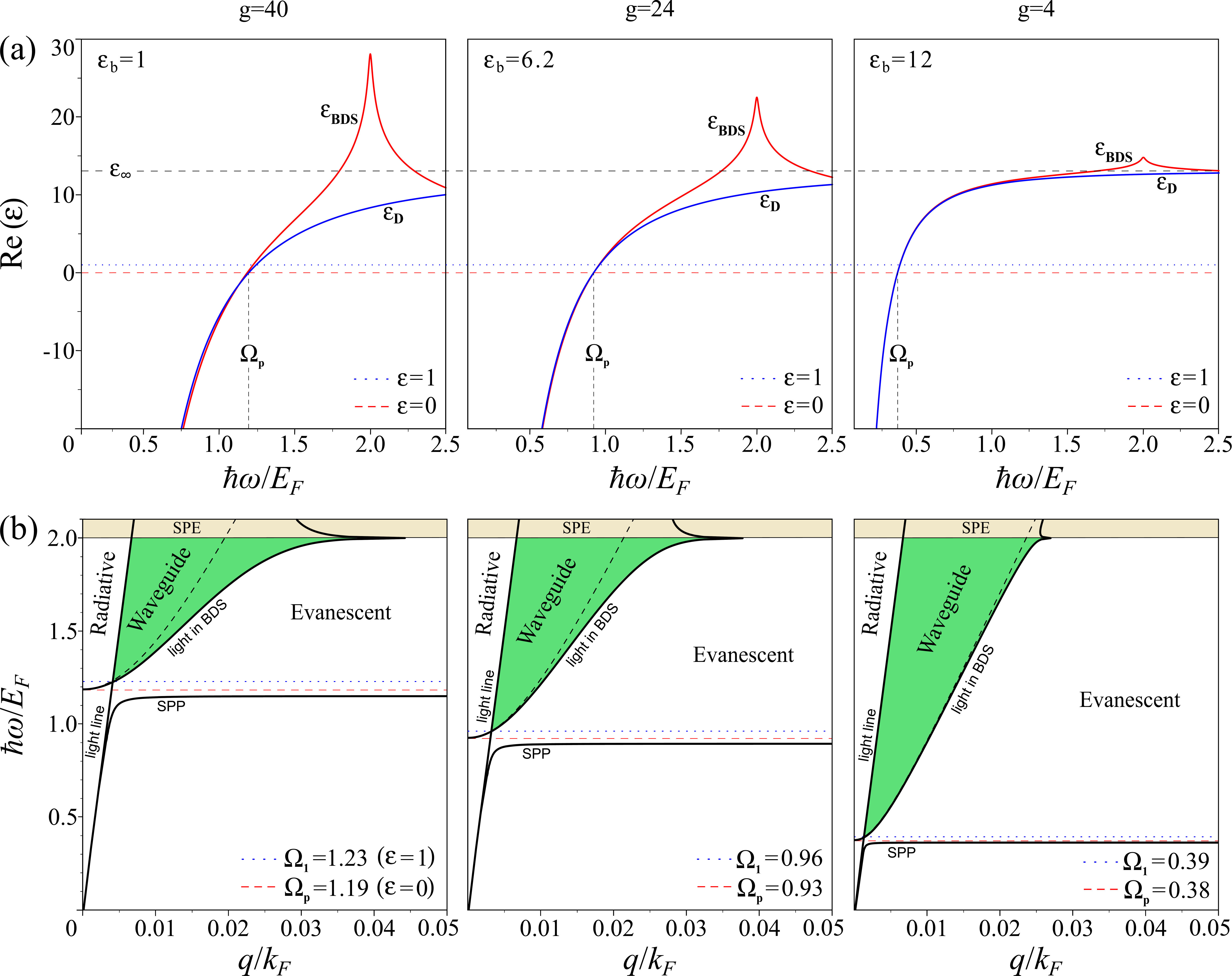}
	\caption{\label{Fig2}(Color online). The real parts of the BDS dielectric functions (a) and the dispersion of light in BDS (b) according to the one-band (Drude) model $\epsilon_\textrm{\tiny D}$ (\ref{epsD}) and two-band model $\epsilon_\textrm{\tiny BDS}$ (\ref{epsBDS}) for different degeneracy factors $g$. The dielectric response region is above $\textup{Re}\, \epsilon=1$ (dotted line). In (b) the region between the line of light out of BDS and the dispersion curve of light in BDS, where WG modes can exist, is shaded by dark color (green online), the region $\hbar\omega/E_F>2$ corresponds to the interband SPE Landau damping regime, and the dashed curve represents the dispersion of light in BDS according to the Drude model (\ref{epsD}). The dielectric functions are taken with $\epsilon_\infty=13$, which for different $g$ factors gives various $\epsilon_b$ (see the text). Other parameters of BDS are the same as for Fig.~\ref{Fig1}.}
\end{figure*}	

\section{Dielectric function and light in the BDS} \label{Sec3}
For the one-band model the dielectric function in RPA can be expressed through intraband polarization function $P_{\rm intra}\!\left(q,\omega\right)$ as $\epsilon\left(q,\omega\right)=\epsilon_\infty-V_qP_{\rm intra}\!\left(q,\omega \right)$, where $\epsilon_\infty$ is the effective background dielectric constant taking into account also interband contributions (usually taken from an experiment as a dielectric constant at infinite frequency) and $V_q =4\pi e^2/q^2$ is the Fourier transform of the bare 3D Coulomb interaction. Alternatively, it can be expressed through the dynamic conductivity: $\epsilon\left(q,\omega\right)=\epsilon_\infty+{4\pi i\sigma_{\rm intra}\!\left(q,\omega \right)}/\omega$. For the BDS case using the first term (the intraband part) of Eq.~(\ref{Imsig}) we obtain the same Drude-like result as in Ref.~\cite{DasSarma_SPP}:  
\begin{equation}
\textup{Re}\, \epsilon_\textrm{\tiny D}\! \left(\Omega\right)=\epsilon_\infty\left(1-\Omega_p^2\big/\Omega^2\right)\label{epsD}
\end{equation} 
where $\Omega_p^2=2r_sg\big/\!\left(3\pi\epsilon_\infty\right)$ is the bulk plasma frequency constant with $r_s =e^2\!\big/\hbar v_F$ being the effective fine structure constant of BDS. This one-band model seems to be enough for considering the metallic response and the behavior of SPPs in BDSs \cite{DasSarma_SPP}. It also roughly characterizes the dielectric response by means of the constant $\epsilon_\infty$, but to describe it more accurately one should use the two-band model taking into account the interband electronic transitions (as was done for the dynamic conductivity in Sec.~\ref{Sec2}). In this model the dielectric function in RPA will be expressed through total polarization function $P=P_{\rm intra}+P_{\rm inter}$ as $\epsilon\left(q,\omega\right)=\epsilon_b-V_qP\left(q,\omega \right)$, where $\epsilon_b$ is the effective background dielectric constant taking into account the contributions from all bands below the valence one. Through the total dynamic conductivity $\sigma=\sigma_{\rm intra}+\sigma_{\rm inter}$ this can be written as: 
\begin{equation}
\epsilon\left(q,\omega\right)=\epsilon_b+{4\pi i\sigma\left(q,\omega \right)}/\omega. \label{eps}          
\end{equation} 
At $q\ll k_F$ and $T\ll E_F$ using Eq.~(\ref{Imsig}) we have:
\begin{equation}
\textup{Re}\, \epsilon_\textrm{\tiny BDS}\! \left(\Omega\right)=\epsilon_b-\frac{2r_sg}{3\pi}\frac{1}{\Omega^2} +\frac{r_sg}{6\pi}\ln\left(\frac{4\varepsilon_c^2}{\left|\Omega^2-4\right|}\right) \label{epsBDS}
\end{equation}            
Finding zeros of Eq.~(\ref{epsBDS}) we obtain the implicit expression for the bulk plasma frequency $\Omega_p$ in BDSs according to the two-band model:
\vskip -15pt
\begin{equation} 
\Omega_p =\sqrt{\frac{2r_sg}{3\pi}\biggl{/}\left(\epsilon_b+\frac{r_sg}{6\pi}\ln\left(\frac{4\varepsilon_c^2}{\left|\Omega_p^2-4\right|}\right)\right)}, \label{w_pl}
\end{equation} 
which coincides with the results obtained in Ref.~\cite{LvZhang} (for BDSs) and Ref.~\cite{ZhouChangXiao} (for WSs with $g=2g_W$, where $g_W$ is the number of pairs of the Weyl nodes). In optical experiments usually the data are fitted with the Drude model giving the constant $\epsilon_\infty$. As the plasma frequencies from the Drude formula (\ref{epsD}) and from the two-band model (\ref{w_pl}) should coincide, we can express $\epsilon_b$ through $\epsilon_\infty$: $\epsilon_b=\epsilon_\infty-\frac{r_sg}{6\pi}\ln\left(\frac{4\varepsilon_c^2}{\left|\Omega_p^2-4\right|}\right)$. Taking $\varepsilon _c=3$  (Ref.~\cite{Ec}),  $\epsilon_\infty=13$ (Ref.~\cite{Sushkov_opt_exp}) for different BDS realizations with various degeneracy factors we obtain the following constants $\epsilon_b$: $\epsilon_b=1$ for $g=40$ (AlCuFe quasicrystals \cite{Basov_opt_exp}), $\epsilon_b=6.2$ for $g=24$ (pyrochlore iridates, e.g., $\textrm{Eu}_2\textrm{IrO}_7$ \cite{Sushkov_opt_exp} or TaAs family \cite{Weng_PRX}), $\epsilon_b=12$ for $g=4$ (including spin degeneracy in $\textrm{Na}_3\textrm{Bi}$ \cite{Liu:Science} or $\textrm{Cd}_3\textrm{As}_2$ \cite{Borisenko,Neupane,Liu:Nature}). In Fig.~\ref{Fig2}(a) for different degeneracy factors we compare the real parts of the BDS dielectric functions according to the one-band (Drude) model (\ref{epsD}) and two-band model (\ref{epsBDS}).

Here we would like to emphasize that the dielectric response does not qualitatively change the plasmon dispersion defined by the equation $\epsilon \left(q,\omega\right)=0$, but influences only the value of the plasma frequency constant $\Omega_p\!\sim\!1/\sqrt{\epsilon_\infty}\, $, whereas for light in a medium one has another type of the governing equation $\epsilon\left(q,\omega\right)=\left(qc/\omega\right)^{2}$, where $c$ is the velocity of light and $q$ is the longitudinal wave vector. In this case the dielectric response can play a crucial role. Indeed, when $\epsilon \left(q,\omega\right)>1$, there can exist the short-wavelength light with $\omega<qc$ like in a dielectric. That is what leads to the existence of the EM modes inside BDS films which will be considered in Sec.~\ref{Sec4}. 
In Fig.~\ref{Fig2}(b) for the same $g$ factors as in Fig.~\ref{Fig2}(a) using Eq.~(\ref{epsBDS}) we plot the dispersion of light in BDS defined by the relation $\Omega^2\epsilon_\textrm{\tiny BDS}\left(\Omega\right)=\left(q/k_F\cdot c/v_F\right )^2$. The dispersion curve of light in BDS starts from $\textup{Re}\, \epsilon=0$ at $\Omega=\Omega_p$ and crosses the dispersion line of light out of BDS when $\textup{Re}\, \epsilon=1$ at $\Omega=\Omega_1$. To the left from the light line $\omega>qc$ and $\omega>{qc/\sqrt\epsilon}$, hence there can be only radiative modes propagating in all directions with the transverse wave vectors $k_\textrm{air}=\sqrt{\left(\omega/c\right)^2 -q^2}$ and $k_\textrm{\tiny BDS}=\sqrt{\epsilon \left(\omega/c\right)^2 -q^2}$ in the media out of BDS and in BDS, respectively. To the right from the dispersion curve of light in BDS $\omega<qc$ and $\omega <{qc/\sqrt\epsilon}$, hence in both media the transverse wave vectors become imaginary and there will be modes evanescent in the transverse direction and propagating in the longitudinal one. Nevertheless, between the light line and the dispersion curve of light in BDS $\omega<qc$ but $\omega >{qc/\sqrt\epsilon}$, therefore only $k_\textrm{air}$ becomes imaginary, which leads to the modes evanescent in the transverse direction out of BDS and propagating in all directions in BDS. That is, in this region [shaded by dark color (green online) in Fig.~\ref{Fig2}(b)] the WG modes can exist. However, at $\Omega>2$ all modes damp due to the interband absorption defined by Eq.~(\ref{Resig}), so this region corresponds to the interband single-particle excitation (SPE) Landau damping regime. Thus at the frequencies $\Omega_1<\Omega<2$, where the dielectric response is dominated, light can penetrate inside BDS and it behaves as a dielectric WG. Notice that the Drude model (\ref{epsD}) also gives the WG region [dashed curve in Fig.~\ref{Fig2}(b)], but at high $g$ factors it is significantly less than the region obtained from the two-band model, when the logarithmic frequency dependence in Eq.~(\ref{epsBDS}) becomes important. Also remark that for $\epsilon_b=1$ (at $g=40$) as seen from Eq.~(\ref{eps}) the dielectric response starts from that frequency $\Omega_1$ [$\textup{Re}\, \epsilon\left(\Omega_1\right)=1$] at which the imaginary part of the conductivity becomes zero $\textup{Im}\, \sigma\left(\Omega_0\right)=0$, i.e., $\Omega_1=\Omega_0=1.23$ (see Sec.~\ref{Sec2}). For other $\epsilon_b$ the frequency $\Omega_1$ may sufficiently differ from $\Omega_0$ [(see Fig.~\ref{Fig2}(b)]. Thus we obtain that the  the dielectric response allows light to penetrate inside BDSs in some ranges of frequencies and wave vectors, but in order to understand what particular WG modes can be excited in BDS films, one should analyse all possible solutions of electrodynamics equations for the system. 

\begin{figure}[t]
	\centering
	\includegraphics[width=0.92\columnwidth]{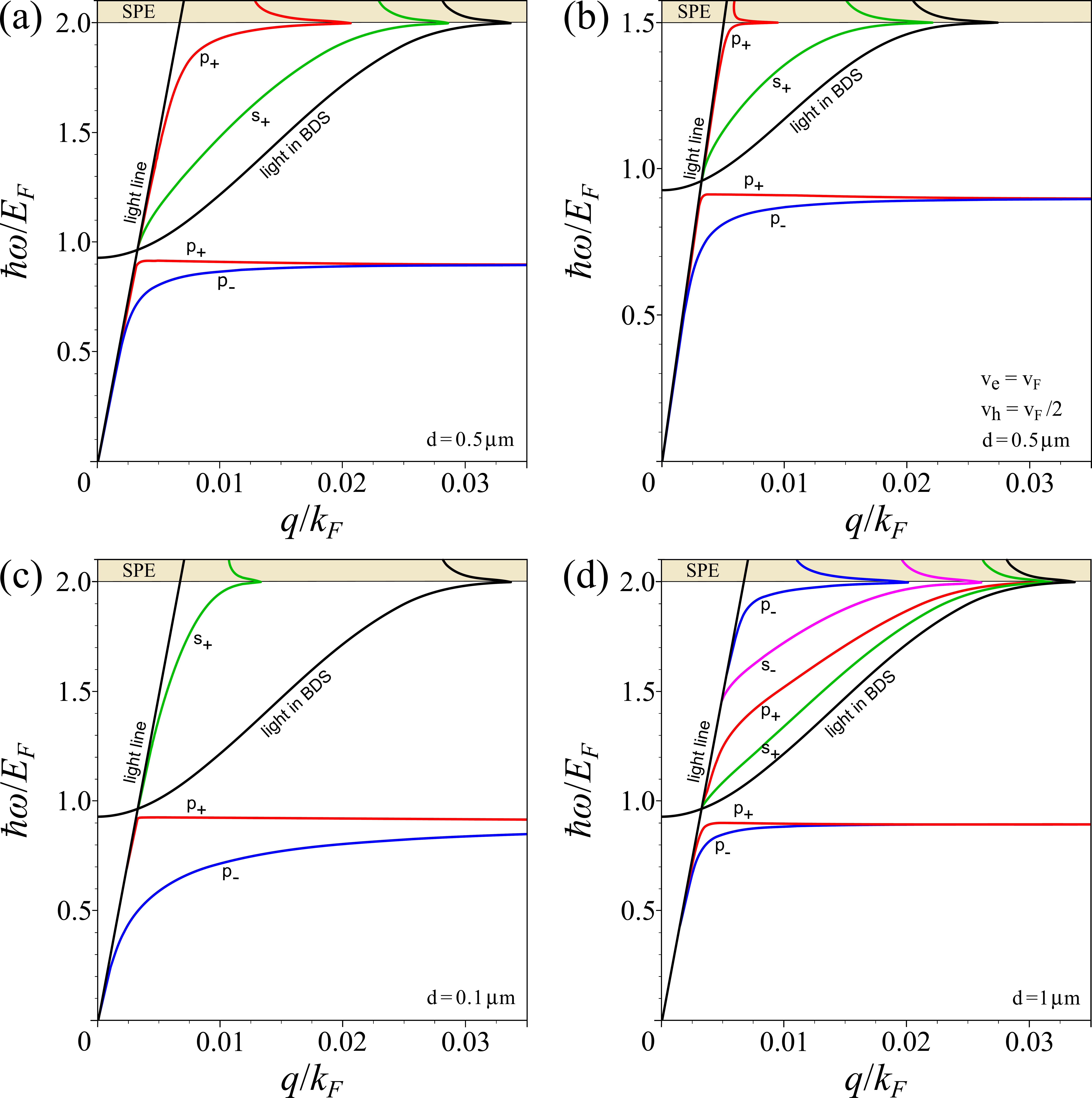} 
	\caption{\label{Fig3}(Color online). The dispersion of the TM and TE waves in BDS films with different thickness $d=0.5\,\mu m$ (a), $d=0.1\,\mu m$ (c), and $d=1\,\mu m$ (d). (b) -- The same as in (a) considering the e-h asymmetry of the Dirac spectrum. The BDS dielectric function (\ref{epsBDS}) is taken with $g=24$, $\epsilon_b=6.2$; other parameters of BDS are the same as for Fig.~\ref{Fig1}. The region SPE corresponds to the interband Landau damping regime.}
\end{figure}

\begin{figure}[t]
	\centering
	\includegraphics[width=1\columnwidth]{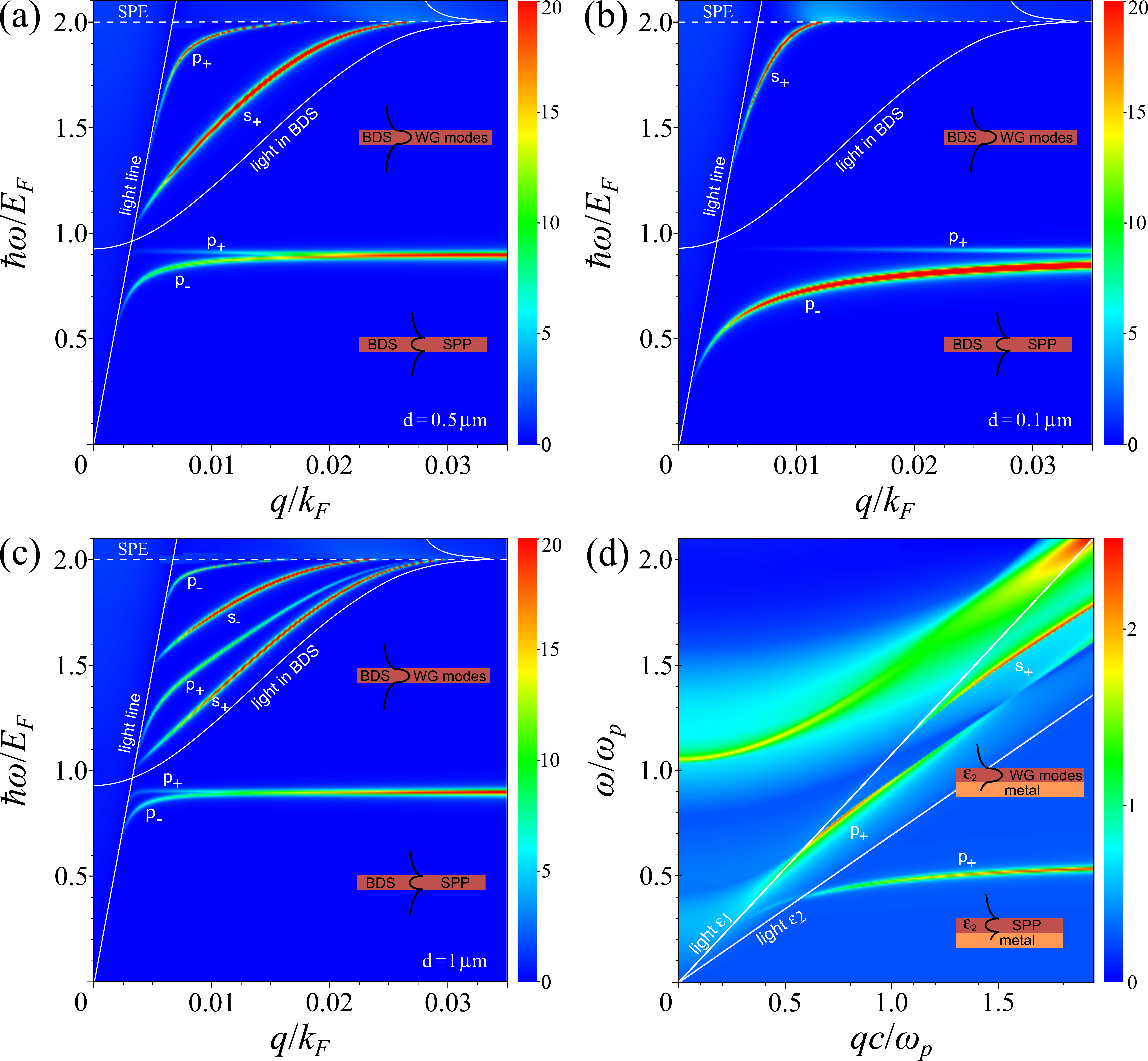}
	\caption{\label{Fig4}(Color online).The loss function (a.u.) of the TM and TE waves in BDS films with different thickness from (a) to (c): $d=0.5,\, 0.1,\, 1\,\mu m$. The parameters of BDS are the same as for Fig.~\ref{Fig3}. The region SPE corresponds to the interband Landau damping regime. (d) -- The loss function (a.u.) of the TM and TE waves in the traditional metal-dielectric waveguide.}
\end{figure}

\section{SPP and EM waves in BDS films} \label{Sec4}

Here we consider all possible solutions for the plane EM waves propagating along a BDS film in the symmetric or asymmetric environment. We calculate dispersion laws, waves field confinement, and loss functions. We also take into account the role of temperature and the influence of the electron-hole asymmetry of the Dirac spectrum.

\subsection{The symmetric environment} \label{Sec4a}

The solution of the electrodynamics equations for the symmetric layer system [a film with the thickness $d$, the dielectric function $\epsilon$, and the transverse wave vector $k_2=\sqrt{q^2-\epsilon\left(\omega/c\right)^2}$ in the environment with $\epsilon_a=1$ and the transverse wave vector $k_1=\sqrt{q^2-\left(\omega/c\right)^2}$] yields the following EM waves dispersion relations \cite{Raether}: 
\begin{align}
\frac{1}{k_1} +\frac{\epsilon}{k_2}\tanh\left(k_2d/2\right)=&\,0\quad(p^-) \label{p-}
\\
\frac{1}{k_1} +\frac{\epsilon}{k_2}\coth\left(k_2d/2\right)=&\,0\quad(p^+) \label{p+}
\end{align} 
for the TM ($p$)-polarized low-frequency mode with the symmetric electric-field profile (\ref{p-}) and the high-frequency mode with the antisymmetric electric-field profile (\ref{p+}). For the TE ($s$) polarization we have:
\begin{align}
k_1 +k_2\tanh\left(k_2d/2\right)=&\,0\quad(s^+) \label{s+}
\\
k_1 +k_2\coth\left(k_2d/2\right)=&\,0\quad(s^-) \label{s-}
\end{align} 
where $(s^+)$ is the high-frequency mode with the antisymmetric electric-field profile (\ref{s+}) and $(s^-)$ is the low-frequency mode with the symmetric electric-field profile (\ref{s-}). Notice that the dispersion relations (\ref{p-}) and (\ref{p+}) for the TM waves (also called SPPs) in the unretarded limit $q\gg\sqrt\epsilon\,\omega/c$ reduced to the in-phase and out-of-phase plasmon dispersion relations, respectively. For thin films at $k_2d\ll1$ using Eq.~(\ref{eps}) we obtain that the dispersion of $p^+$ and $s^-$ (``coth modes'') degenerates ($p^+$ reduces to $\omega =\omega_p$ and $s^-$ do not exist) and the dispersion of  $p^-$ and $s^+$ (``tanh modes'') in thin films with the 3D dynamic conductivity $\sigma_\textrm{\tiny 3D}$ will be the following: 
\begin{align}
&1\big/k_1\approx -2\pi i\sigma_\textrm{\tiny 3D}d\big/\omega\quad(p^-) \label{thin_p}
\\
&k_1\approx 2\pi i\sigma_\textrm{\tiny 3D}d\omega\big/c^2\quad(s^+) \label{thin_s}
\end{align}
for the symmetric TM $(p^-)$ (\ref{thin_p}) and the antisymmetric TE $(s^+)$ (\ref{thin_s}) waves. This corresponds to the EM waves dispersion relations in the 2D electron gas (2DEG) systems (e.g., graphene) (see Refs.~\cite{Stern,Nakayama}) with the 2D dynamic conductivity $\sigma_\textrm{\tiny 2D}=\sigma_\textrm{\tiny 3D}d$. As was mentioned above, graphene possesses both the TM waves (at low frequencies when $\textup{Im}\, \sigma>0$) and the TE waves (at frequencies when $\textup{Im}\, \sigma <0$). Hence, due to the similar behavior of the BDS conductivity (see Sec.~\ref{Sec2}), BDS films can support not only SPPs (the TM waves), but also the TE waves inside the film, a 3D analog of the TE waves in graphene. These waves are the WG modes, the manifestation of the dielectric response in BDSs (see Sec.~\ref{Sec3}).

Substituting Eq.~(\ref{epsBDS}) in Eqs.~(\ref{p-})--(\ref{s-}) we obtain the dispersion laws (Fig.~\ref{Fig3}) and the loss functions (Fig.~\ref{Fig4}) of the TM and TE waves in BDS films with the different thicknesses $d$. The loss function of EM waves with the dispersion equation $f\left(q,\omega \right)=0$ determines the measure of the wave damping and can be defined by $\textup{-Im}\, \left[f\left(q,\omega\right)^{-1}\right]$. The undamped waves (the solution for both $\textup{Re}\, f$ and $\textup{Im}\, f$ becomes zero) displayed in the loss function as a well defined $\delta$-function peak. Thus the measure of the wave damping is expressed by the broadening of the peak in the loss function -- if the wave is overdamped, there will be no peak in the loss function. At $d=0.5\,\mu m$ (see Fig.~\ref{Fig3}(a)) we obtain not only the symmetric $(p^-)$ and the antisymmetric $(p^+)$ SPP modes, but also the TM-polarized $(p^+)$ and the TE-polarized $(s^+)$ antisymmetric WG modes. Fig.~\ref{Fig4}(a)) shows that these WG modes will be not less pronounced than the SPP modes, moreover, the TE wave $(s^+)$ is even less damped than SPPs. With decreasing of the thickness of the film the high-frequency SPP mode reduces to $\omega =\omega_p$ and the WG modes tend to the light line and become vanishing. At $d=0.1\,\mu m$ [see Figs.~\ref{Fig3}(c) and \ref{Fig4}(b)] among the WG modes only $s^+$ will exist. On the other hand, with the increasing of the thickness the SPP modes merge into one and, in addition to the antisymmetric WG modes, the symmetric TM $(p^-)$ and TE $(s^-)$ WG modes appear. At $d=1\,\mu m$ all these types of WG modes can be observed [see Fig.~\ref{Fig3}(d)], but as seen from Fig.~\ref{Fig4}(c) they will be twice stronger damped. Therefore, with the increasing of the thickness the number of the WG modes grows, but also their damping rises. Thus the optimal thickness of BDS WGs lies in the interval $0.5-1\,\mu m$. For the comparison we have calculated the loss function of the TM and TE waves in the traditional metal-dielectric WG. It also possesses different WG modes in the dispersion region between the light line out of WG (light in $\epsilon_1$) and the light line in the dielectric layer (light in $\epsilon_2$) [see Fig.~\ref{Fig4}(d)]. But unlike in BDS WGs, here the WG region starts from the zero frequency and its boundaries have the linear dispersion. The main advantage of a BDS WG over a metal-dielectric one is that it consists of a single material, but supports both SPP and WG modes at the corresponding frequencies.

According to the experimental data \cite{Borisenko,Neupane,Liu:Nature} some BDSs have a significant e-h asymmetry of the Dirac spectrum. As we have shown in Appendix \ref{Appendix_B} the contribution of this asymmetry to the conductivity can be accounted for by the factor $\gamma=\left(v_+/v_-+1\right)\! /2$, where $v_-$ and $v_+$ are the velocities of electrons and holes, respectively. For the realistic parameters $v_-\equiv v_F$, $v_+ =v_-/2$ (see, e.g., Ref.~\cite{Borisenko}) the factor is $\gamma=3/4$, which causes the shift of the interband damping region SPE: as seen from Eq.~(\ref{Resig_B}) the damping region starts from $\Omega=2\gamma=1.5$ instead of $\Omega =2$. Also this asymmetry causes the shift with the compression of the WG region [compare Figs.~\ref{Fig3}(a) and \ref{Fig3}(b)], which can suppress the TM WG mode $(p^+)$. 

\begin{figure}[t]
	\centering
	\includegraphics[width=1\columnwidth]{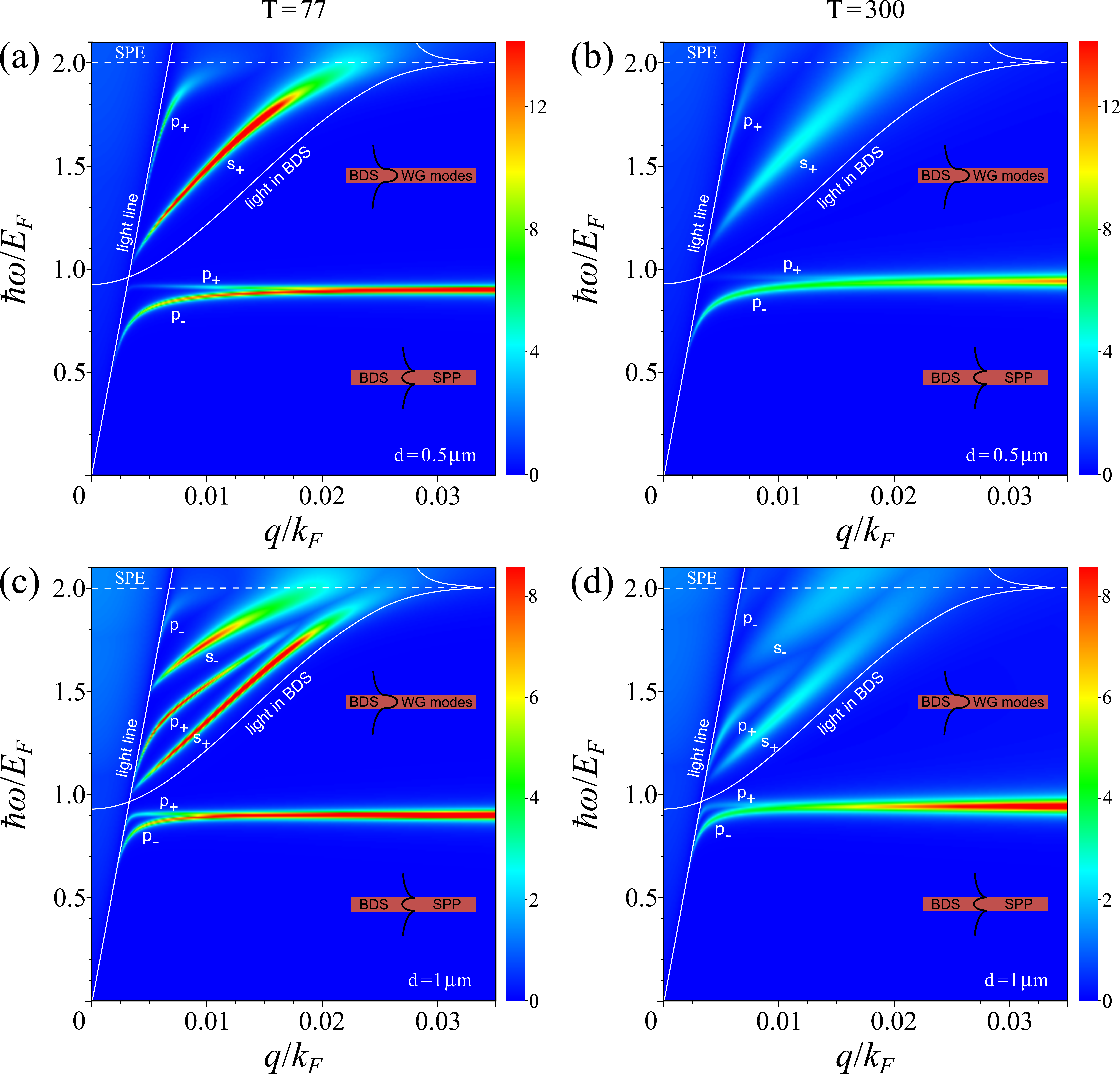} 
	\caption{\label{Fig5}(Color online). The loss function (units as in Fig.~\ref{Fig4}) of the TM and TE waves in a BDS film  with thickness $d=0.5\,\mu m$ (a), (b) and $d=1\,\mu m$ (c), (d) at nitrogen temperature $T=77$K (a), (c) and at room temperature $T=300$K (b), (d). Other parameters of BDS are the same as in Fig.~\ref{Fig3}. The region SPE corresponds to the interband Landau damping regime.}
\end{figure}

Calculating numerically the integral in Eq.~(\ref{ImsigT}) and neglecting the temperature dependence of the mobility we compare the loss function of the WG modes and SPP in BDS films at low temperature (we take 77 K) and at room temperature (300 K). As seen from Fig.~\ref{Fig5} temperature does not greatly affect SPP  but suppresses the WG modes, though not destroying them. For $d=0.5\,\mu m$ at $T=300$K only TE mode $s^+$ will survive [see Fig.~\ref{Fig5}(b)] and for $d=1\,\mu m$ at $T=77$K all modes except $p^-$ exist, but at $T=300$K only $p^+$ and $s^+$ are still pronounced [see Fig.~\ref{Fig5}(d)]. In any case among all WG modes the TE wave $s^+$ is the strongest one.

\begin{figure}[t]
	\centering
	\includegraphics[width=0.7\columnwidth]{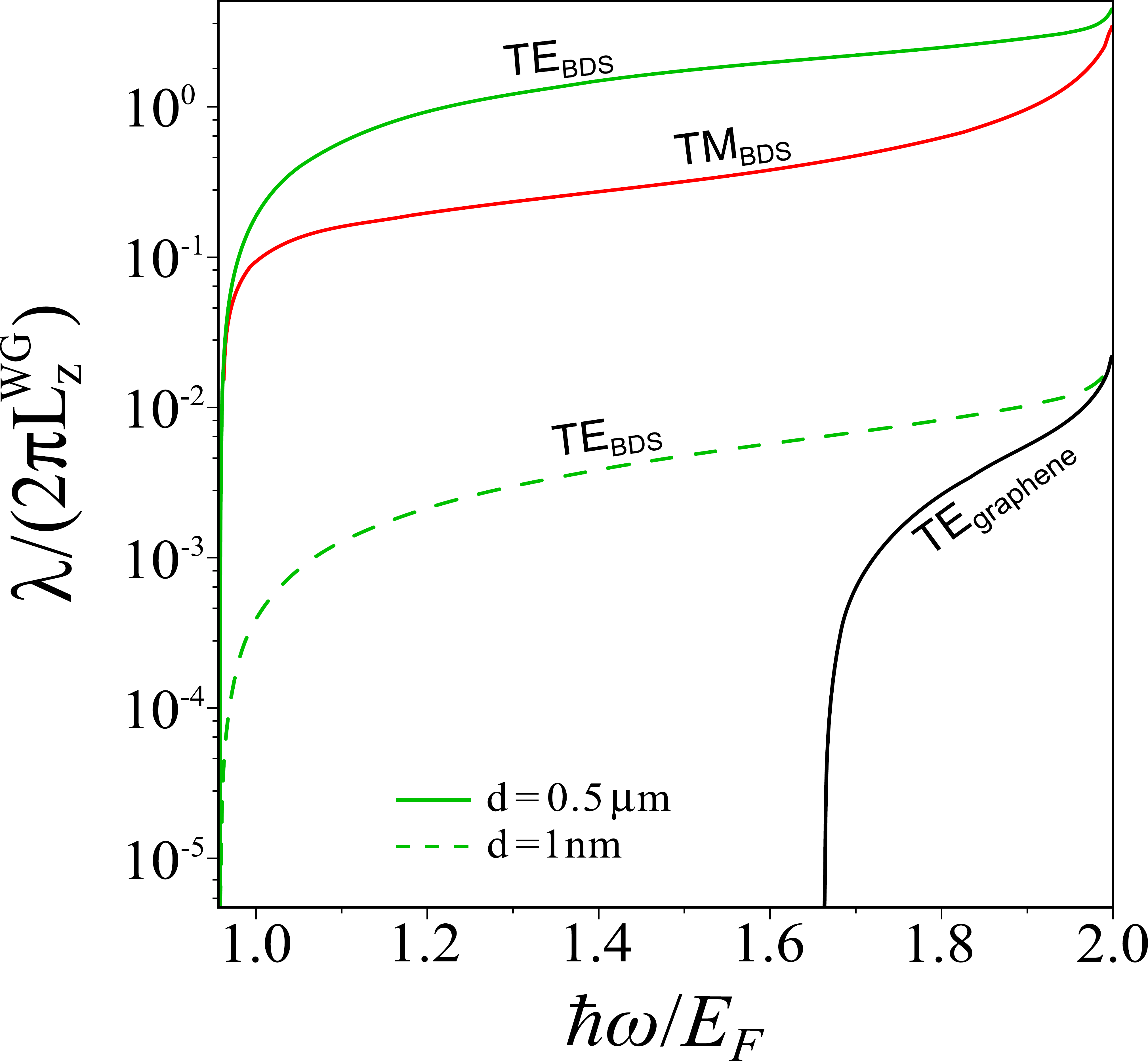} 
	\caption{\label{Fig6}(Color online). The field confinement factor of the WG modes in a BDS film with thickness $d=0.5\,\mu m$ (solid colored), $d=1\, $nm (dashed) and the confinement factor for the TE waves in graphene (solid black). The parameters of BDS are the same as in Fig.~\ref{Fig3} and for graphene as in Fig.~\ref{Fig1}.}
\end{figure}

We also calculated the field confinement factor $\lambda\big/2\pi L_{z}^{WG}$ of the WG modes, defined by the ratio of the free-space-light wavelength $\lambda$ and the WG modes decay length (in the direction transverse to the film) $L_{z}^{WG}=1/\left|k_{1z}\right|=1\! \bigg/\! \sqrt{q^2-\left(\omega/c\right)^2}$ corresponding to the $1/e$ field decay. This factor indicates the measure of how strongly the WG modes are pinned to the film surface. The confinement factor of the WG modes in BDS films decreases with the reducing of the thickness: at $d=0.5\,\mu m$ it is by two orders of magnitude higher than at $d=1\, $nm (see Fig.~\ref{Fig6}). Comparing with graphene in a free space with the dispersion of the TE waves given by $k_1=2\pi i\sigma_\textrm{\tiny Gr}\omega\big/c^2$, from Fig.~\ref{Fig6} one can see that in a BDS film with the thickness $d\geq1\, $nm the TE WG modes will be pinned to the surface of the film greater than the evanescent TE waves pressed to graphene. Thickness reduction of a BDS film up to the atomic layer (other words in the case of the 3D-2D Dirac spectrum crossover) will lead to the vanishing of the WG TM mode (see Fig.~\ref{Fig6})  and to the conversion of the WG TE mode to the evanescent graphene-like TE wave [see Eq.~(\ref{thin_s}), where $\sigma_\textrm{\tiny 3D}=\sigma_\textrm{\tiny Gr}/d$]. Notice that for the TM waves the decay length is proportional to the conductivity $L_z=1/\!\left|k_z\right|\sim\ \!\!\!\left|\sigma\right|$, while for the TE waves the reverse situation takes place $L_z=1/\!\left|k_z\right|\sim\ \!\!1/\!\left|\sigma\right|$.
 
\subsection{The asymmetric environment} \label{Sec4b}

To solve this problem one should consider the solutions of the electrodynamics equations for the asymmetric layer system: the film with the thickness $d$, the dielectric function $\epsilon$, and the transverse wave vector $k_2=\sqrt{q^2-\epsilon\left(\omega/c\right)^2}$; the medium above the film with $\epsilon_1=1$ and the transverse wave vector $k_1=\sqrt{q^2-\epsilon_1\left(\omega/c\right)^2}$; and the medium under the film (the semi-infinite substrate) with $\epsilon_3$ and the transverse wave vector $k_3=\sqrt{q^2-\epsilon_3\left(\omega/c\right)^2}$. For the TM waves $\left(p^{\pm}\right)$ we have: 
\begin{equation}  
\left(\frac{k_1k_3}{\epsilon_1\epsilon_3}+\frac{k_2^2}{\epsilon^2}\right)\tanh\left(k_2 d\right)+\left(\frac{k_1}{\epsilon_1} +\frac{k_3}{\epsilon_3}\right)\frac{k_2}{\epsilon} =0 \label{subs_p}
\end{equation} 
and for the TE waves $\left(s^{\pm}\right)$:
\begin{equation} 
\left(k_1k_3+k_2^2\right)\tanh\left(k_2d\right)+\left(k_1+k_3\right)k_2=0 \label{subs_s}
\end{equation} 

\begin{figure}[t]
	\centering
	\includegraphics[width=1\columnwidth]{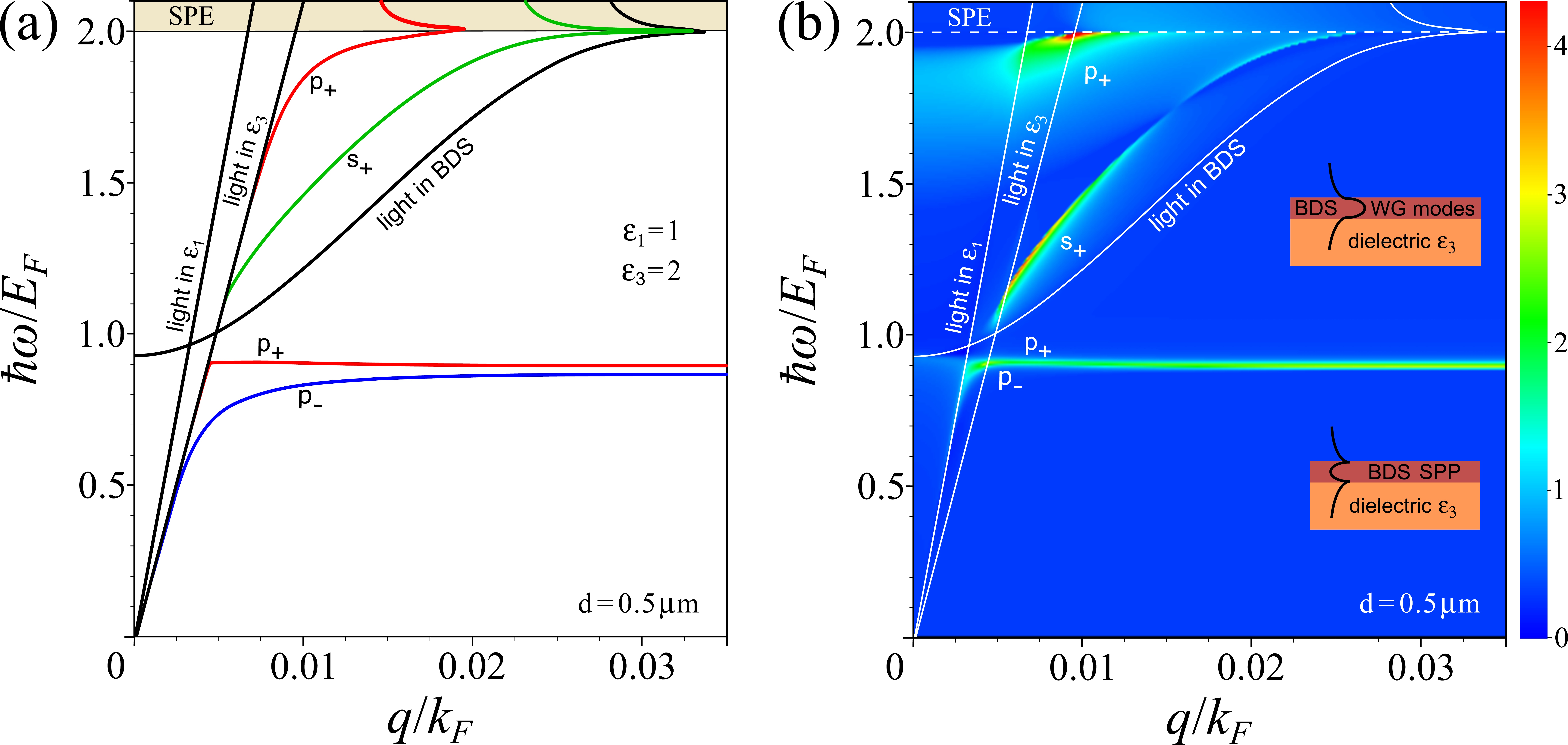} 
	\caption{\label{Fig7}(Color online). The dispersion (a) and the loss function (a.u.) (b) of the TM and TE waves in a BDS film with thickness $d=0.5\,\mu m$ in the asymmetric environment: $\epsilon_1=1$, $\epsilon_3=2$ (substrate). Other parameters of BDS are the same as in Fig.~\ref{Fig3}. The region SPE corresponds to the interband Landau damping regime.}
\end{figure}

Taking the $\textrm{SiO}_2$ substrate with $\epsilon_3=2$ (for the frequencies in the WG region $0.14-0.3$eV) we get that the WG modes do not exist between the dispersion lines of light in $\epsilon _1$ and in $\epsilon_3$  (they leak into the substrate $\epsilon_3$) and exist only in the region between the dispersion line of light in $\epsilon_3$ and the dispersion curve of light in a BDS [see Fig.~\ref{Fig7}(a)]. Moreover, in this region they are sufficiently suppressed [compare Fig.~\ref{Fig7}(b) and \ref{Fig4}(a)]. For the dielectric constant of the substrate larger than $\epsilon_3\approx15$ the dispersion curve of light in BDS lies in the cone of light in $\epsilon_3$ and hence all WG modes become leaky and do not propagate along a BDS film. Notice that the same effect takes place in the case of the symmetric environment (see Sec.~\ref{Sec4a}) with $\epsilon_a\geq15$. Therefore, to avoid the waves leakage BDS WGs should be placed on the low-$\epsilon$ substrates \cite{Low-n} or just suspended.


\begin{figure*}[t]
	\centering
	\includegraphics[scale=0.115]{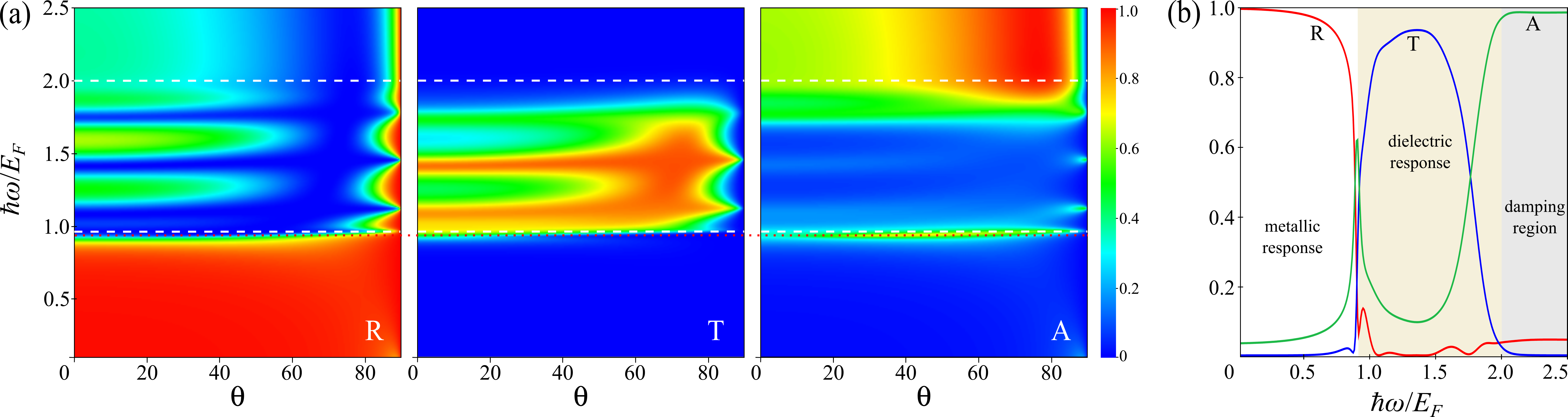} 
	\caption{\label{Fig8}(Color online). (a) -- The reflection (R), transmission (T) and absorption (A) energy spectra of the TM-polarized light incident on the BDS film with the thickness $d=2\,\mu m$ surrounded by the medium with $\epsilon_1=1$ at $T=77$K versus normalized frequency and incidence angle. The dotted red line displays the bulk plasma frequency and the dashed white lines display the frequency window of the dielectric response with weak damping. (b) -- The cross section of the RTA spectra at $\theta=70^{\circ}$. Other parameters of BDS are the same as in Fig.~\ref{Fig3}.}
\end{figure*}

\section{Optical spectra of BDS film} \label{Sec5}

In this section we consider the influence of the dielectric response in BDSs on the optical spectra of light incident on a BDS film. The reflection (R), transmission (T) and absorption (A) energy coefficients for the nonmagnetic layer with the thickness $d$, the refractive index $n_2=\sqrt{\epsilon_2}$, and the transverse wave vector $k_2=\omega/c\cdot n_2\! \cos\theta_2$ in the environment with $n_1 =\sqrt{\epsilon_1}$ and the transverse wave vector $k_1=\omega/c\cdot\! n_1\!\cos\theta_1$ are expressed by \cite{Born-Wolf}: 
\begin{align}\nonumber
\textrm{R}=&\,\left|\frac{r_{12}\left(1-\exp \left(2ik_2d\right)\right)}{1-\left(r_{12}\right)^2\exp \left(2ik_2d\right)}\right|^2,
\\[0.5em]\nonumber
\textrm{T}=&\,\left|\frac{k_2}{k_1} \frac{\left(t_{12}\right)^{2}\exp\left(ik_2d\right)}{1-\left(r_{12}\right)^{2}\exp\left(2ik_2d\right)}\right|^2,  
\\[0.5em]
\textrm{A}=&\,1-\textrm{R}-\textrm{T}, \label{RTA}
\end{align}
where the Fresnel coefficients different for each polarization are:
\begin{align}\nonumber
    r_{12}^{TE} =\frac{k_1 -k_2}{k_1+k_2},&\quad r_{12}^{TM} =\frac{k_1/\epsilon_1 -k_2/\epsilon_2}{k_1/\epsilon_1+k_2/\epsilon_2}
    \\[0.5em]
   t_{12}^{TE} =\frac{2k_1}{k_1+k_2},&\quad t_{12}^{TM}=\frac{2k_1/\sqrt{\epsilon_1\epsilon_2}}{k_1/\epsilon_1+k_2/\epsilon_2}. \label{Fresnel}
\end{align} 
Substituting Eq.~(\ref{epsBDS}) in Eqs.~(\ref{RTA}) we obtain RTA spectra of the TM-polarized light incident on a BDS film. As seen from Fig.~\ref{Fig8} the BDS shows the typical metallic behavior but in the mid-IR region: the absorption peak at the bulk plasma frequency and the total reflection at lower frequencies. However, unlike in metal, in BDSs the dielectric response arises at $\Omega>0.96$ (see Sec.~\ref{Sec3}), which causes: the typical dielectric films oscillations in the reflection, the wide-angle passband in the frequency window $\Omega\!\in\!\left[0.96,2\right]$ in the transmission, the wide-angle plasmon absorption peak, and the total absorption region at $\Omega>2$ corresponding to the interband electronic transitions in BDS. The passband is limited from the bottom by the reflection edge (represented by the the plasma frequency) and from the top by the total absorption edge: $\Omega\!\in\!\left[\Omega_p,2E_F\right]$. For the typical Fermi level $E_F=0.15$eV the frequency window of the wide-angle passband in a BDS film lies in the mid-IR range ($\lambda\!\in\!\left[4.1,8.6\right]\mu m$), which can be used for the omnidirectional mid-IR transmission filtering. The deviation of the passband and the total absorption region from the edge $\Omega=2$ (see Fig.~\ref{Fig8}) is connected with the temperature smearing. Notice that the region of transparency appears in some semiconductors or even metals \cite{CorrMetals}. Widely used ITO and ZnO, being highly doped semiconductors with the plasma frequency $w_p\approx1$eV, have large electronic band gap $\Delta\approx3.75$eV resulting in rather broad photonic passband $\Omega\!\in\!\left[\omega_p,2E_F+\Delta\right]\approx[1,4]$eV. Moreover, they possess very low carrier mobility, less than $60\mathrm{cm^2V^{-1}s^{-1}}$ (Ref.~\cite{Ellmer}). The Dirac nature of the electron spectrum in BDSs causes two main benefits over common transparent conductors: zero electronic band gap giving the narrow passband and symmetry protection that strongly suppresses backscattering, which results in the ultrahigh mobility (see Sec.~\ref{Sec1}). In addition notice that unlike Fabry-Perot omnidirectional bandpass filters based on the interference effects (see, e.g., Ref.~\cite{Park}), in the BDS filter the passband is a result of the BDS electronic properties. Being interferenceless the BDS filter can be used for geometrically independent filterring. 

\section{Conclusion} \label{Sec6}

Using the Kubo formalism in RPA we have calculated the BDS local dynamic conductivity and the dielectric function and found that at frequencies lower than Fermi energy the metallic response in a BDS film manifests in the existence of SPP, but at higher frequencies the dielectric response is dominated and a BDS film behaves as a dielectric WG. At this dielectric regime we predict the existence of novel TM- and TE-polarized EM modes propagating in BDS WGs, a 3D analog of the TE waves in graphene. However, this WG modes at room temperature will be rather suppressed, though still exist (mainly the TE mode). Besides, they strongly leak into a substrate, thus BDS WGs should be placed on the low-$\epsilon$ substrates or suspended. We estimate that the optimal thickness of BDS WGs lies in the interval $0.5-1\,\mu m$. With an increase of the thickness the additional sets of modes will appear, but their spectral strength will reduce. With a decrease of the thickness the WG TM mode will disappear and the WG TE mode will convert to the graphene-like evanescent TE wave. We also find that the dielectric response manifests as the wide-angle passband in the mid-IR transmission spectrum of light incident on a BDS film, which can be used for the omnidirectional mid-IR bandpass filtering. Moreover, being interferenceless the BDS filter could provide unique opportunities for geometrically independent filterring. The tuning of the Fermi level of the system allows us to switch between the metallic and the dielectric regimes and to change the frequency range of the predicted WG modes. All this makes BDSs promising materials for photonics and plasmonics. 

\section*{Acknowledgments}

The authors are grateful to \mbox{A. A. Sokolik} and \mbox{E. S. Andrianov} for useful discussions. The work was supported by the Grant of Global Research Outreach in Samsung Advanced Institute of Technology and by Russian Foundation for Basic Research. Yu. E. L. thanks the Basic Research Program of the National Research University Higher School of Economics.

\onecolumngrid
\appendix
\section{Longitudinal local dynamic conductivity of the Dirac 3DEG}
\label{Appendix_A}

The optical response of the Dirac 3DEG with the low-energy spectrum $E_{\mathbf k,s} =s\hbar v_Fk$, where $k$ is the 3D momentum magnitude, $v_F$ is the Fermi velocity of a Dirac fermion, and $s=\pm1$ denote the band indices, is described by the dynamic conductivity tensor. For non-interacting electrons in the local response approximation this tensor can be written in the Kubo-Greenwood formulation as
\begin{equation} 
\sigma_{\alpha\beta}(\omega)=\frac{-ie^2g\hbar}{V}\sum_{\mathbf k,s,s'}\frac{n\left(E_{\mathbf k,s}\right)-n\left(E_{\mathbf k,s'}\right)}{E_{\mathbf k,s} -E_{\mathbf k,s'}}\frac{\left\langle \mathbf ks\right|\widehat v_\alpha\left|\mathbf ks'\right\rangle\left\langle \mathbf ks'\right|\widehat v_\beta\left|\mathbf ks\right\rangle}{\hbar(\omega +i0)+E_{\mathbf k,s}-E_{\mathbf k,s'}}. 
\end{equation}
Here $\alpha =\left(x,y,z\right)$, $\omega$ is the frequency of the incident electromagnetic wave, $V$ is the 3DEG volume, $g$ is the degeneracy factor, and $\widehat v_\alpha=v_F\sigma _\alpha$ is the velocity operator, where $\sigma_\alpha$ are the Pauli matrices, $\left\langle \mathbf ks\right|$ and $\left|\mathbf ks'\right\rangle$ are the initial and the final electron states of the Dirac 3DEG described by the Hamiltonian $\widehat H=\hbar v_F\boldsymbol\sigma\mathbf k$, and $n\left(E_{k,s}\right)=1\bigl/\left(\exp\left(\left(E_{k,s}-E_F\right)/T\right)+1\right)$ is the Fermi distribution function with the Fermi level $E_F$ and temperature $T$ in the energy units. Therefore the intraband and the interband contributions in the longitudinal dynamic conductivity can be expressed as
\begin{align}
\sigma_{xx}^{\rm intra}(\omega)=&\,\frac{-ie^2g}{(\omega+i0)V}\sum_{\mathbf k}\frac{\partial n\left(E_\mathbf k\right)}{\partial E_\mathbf k}v_x^2, \label{sxx_intra}
\\
\sigma_{xx}^{\rm inter}(\omega )=&\,\frac{-ie^2g\hbar}{V}\sum _{\mathbf k,s\ne s'}\frac{n\left(E_{\mathbf k,s}\right)-n\left(E_{\mathbf k,s'}\right)}{E_{\mathbf k,s} -E_{\mathbf k,s'}}\frac{\left|\left\langle \mathbf ks\right|\widehat v_x\left|\mathbf ks'\right\rangle\right|^2}{\hbar(\omega +i0)+E_{\mathbf k,s}-E_{\mathbf k,s'}}.\label{sxx_inter}
\end{align}
In this work we operate only with the longitudinal conductivity and for simplicity omit the subscript: $\sigma_{xx}\equiv\sigma$. The spinor part of the eigenfunctions of the 3D Dirac Hamiltonian, corresponding to an electron with the momentum $\mathbf k$ [defined in the 3D space by the azimuthal ($\varphi$) and the polar ($\theta $) angles] from the conduction $\left(s=-1\right)$and the valence $\left(s=+1\right)$ bands, can be written as
\begin{equation}
\left|\mathbf k+\right\rangle=\binom{\cos(\theta/2)}{\rm e^{i\varphi}\sin(\theta/2)},\quad \left|\mathbf k-\right\rangle=\binom{-\sin(\theta/2)}{\rm e^{i\varphi}\cos(\theta/2)}. \label{k}
\end{equation}
Writing Eq.~(\ref{sxx_intra}) in the integral form we get: 
\begin{equation}
\sigma_{\rm intra}(\omega)=\frac{-ie^2g}{\omega V}\int_{-\infty}^{\infty}\frac{4\pi k^2dk}{\left(2\pi\right)^3\!\!/V}\frac{\partial n(E)}{\partial E}\int_{\Omega_{3\rm D}}\frac{v_x^2}{4\pi},
\end{equation}
where the last one is the integral over the solid angle $\Omega_{3\rm D}$ in the 3D space. Using $k=E/\hbar v_F$ and calculating $\int_{-\infty}^{\infty}E^2\frac{\partial n(E)}{\partial E}dE=-E_F^2-\pi^2T^2\big/3$, $\int_{\Omega_{3\rm D}}\frac{v_x^2}{4\pi}=v_F^2\big/3$ we finally obtain:  
\begin{equation}
\sigma_{\rm intra}(\omega )=\frac{ie^2}{\hbar}\frac{gk_F}{6\pi^2\Omega}\left(1+\frac{\pi^2}{3} \left(\frac{T}{E_F}\right)^2\right), \label{s_intra}
\end{equation}
where $\Omega =\hbar\omega/E_F$, $k_F=E_F/\hbar v_F$ is the Fermi momentum. The interband conductivity (\ref{sxx_inter}) in the integral form will be:
\begin{equation}
\sigma_{\rm inter}(\omega)=\frac{-ie^2g\hbar}{V}\int_{\Omega_{3\rm D}}\frac{\left|\left\langle \mathbf k+\right|\widehat v_x\left|\mathbf k-\right\rangle\right|^2}{4\pi}\int_{0}^{\infty}\frac{4\pi k^2dk}{\left(2\pi\right)^3\!\!/V}
\times\left[\frac{n(E)-n(-E)}{2E}\left(\frac{1}{\hbar(\omega+i0)+2E}+\frac{1}{\hbar(\omega+i0)-2E}\right)\right]. \label{s_inter}
\end{equation}
Using $k=E/\hbar v_F$ and calculating with Eq.~(\ref{k}) $\int_{\Omega_{3\rm D}}\left|\left\langle \mathbf k+\right|v_F\sigma_x\left|\mathbf k-\right\rangle\right|^2\!\big/4\pi=2v_F^2\big/3$ we get: 
\begin{equation}
\sigma_{\rm inter}(\omega)=\frac{-ie^2g\omega}{3\pi^2\hbar v_F}\int_{0}^{\infty}\left(\frac{n(E)-n(-E)}{\hbar^2(\omega+i0)^2-4E^2}\right)EdE.
\end{equation}
As for the 2D case (e.g., graphene \cite{Falkovsky}) one can resolve the singularity $E=\hbar\omega/2$ rewriting the integral in the form useful for numerical calculations:
\begin{equation}
\sigma_{\rm inter}(\omega)=\frac{ie^2g\omega}{3\pi^2\hbar v_F}\left[-\frac{\pi i}{2}\frac{G(\hbar\omega/2)}{4}+\int_{0}^{\infty}\left(\frac{G(E)-G(\hbar\omega/2)}{\hbar^2\omega^2-4E^2}\right) EdE\right],
\end{equation}
where $G(E)=n(-E)-n(E)=\frac{\sinh(E/T)}{\cosh(E_F/T)+\cosh(E/T)}$. Finally, taking into account Eq.~(\ref{s_intra}) we obtain that the real and imaginary parts of the longitudinal dynamic conductivity $\sigma =\sigma_{\rm intra} +\sigma_{\rm inter}$ are expressed as
\begin{align}
\textup{Re}\, \sigma\! \left(\Omega\right )=&\,\frac{e^2}{\hbar}\frac{gk_F}{24\pi}\Omega G\left(\Omega/2\right), \label{ResigT_A}
\\ 
\textup{Im}\, \sigma\! \left(\Omega\right)=&\,\frac{e^2}{\hbar}\frac{gk_F}{24\pi^2}\biggl[\frac{4}{\Omega}\left ( 1+\frac{\pi^2}{3}\left(\frac{T}{E_F}\right)^{2}\right)+8\Omega\int_{0}^{\varepsilon _c}\left (\frac{G\left ( \varepsilon \right )-G\left ( \Omega/2 \right )}{\Omega^2-4\varepsilon ^2}\right )\varepsilon d\varepsilon\biggr], \label{ImsigT_A}
\end{align}	
where $\varepsilon=E/E_F$ and $\varepsilon_c =E_c/E_F$ ($E_c$ is the cutoff of energy: unlike the 2D case, in the 3D case the integral diverges). At the low-temperature limit $kT\ll E_F$ $G(\Omega/2)\to \rm{\theta}(\Omega -2)$, and we obtain: 
\begin{align}
\textup{Re}\, \sigma\! \left(\Omega\right)=&\,\frac{e^2}{\hbar }\frac{gk_F}{24\pi}\Omega\,\rm{\theta}(\Omega-2), \label{Resig_A}
\\
\textup{Im}\, \sigma\! \left(\Omega\right)=&\,\frac{e^2}{\hbar}\frac{gk_F}{24\pi^2}\left[\frac{4}{\Omega }-\Omega\ln\left(\frac{4\varepsilon_c^2}{\left|\Omega^2-4\right|}\right)\right]. \label{Imsig_A}
\end{align}

\section{The case of the electron-hole asymmetry in the 3DEG Dirac spectrum}
\label{Appendix_B}

In the case of the e-h asymmetry of the low-energy Dirac spectrum with $E_{k,s}=sv_s\hbar k$, where $v_s$ is the Fermi velocity different for each band ($v_-$ for electrons and $v_+$ for holes), the intraband conductivity (\ref{s_intra}) remains the same as for the symmetrical case, but the interband one [Eq.~(\ref{s_inter})] must be rewritten as
\begin{align} \nonumber 
\sigma_{\rm inter}(\omega)=\frac{-ie^2g\hbar}{V}\int_{\Omega_{3\rm D}}\frac{\left|\left\langle \mathbf k+\right|\widehat v_x\left|\mathbf k-\right\rangle\right|^2}{4\pi}&\int_{0}^{\infty}\frac{4\pi k^2dk}{\left(2\pi\right)^3\!\!/V}
\\
&\!\!\!\!\!\!\times\left[\frac{n(E_+)-n(E_-)}{E_+-E_-}\left(\frac{1}{\hbar(\omega+i0)+(E_+-E_-)}+\frac{1}{\hbar(\omega+i0)-(E_+-E_-)}\right)\right], \label{s_inter_B} 
\end{align}
where the velocity operator should be defined in the general form $\widehat v_x =\frac{1}{\hbar} \left(\frac{\partial\widehat H}{\partial\mathbf k}\right)_x$. Using the spectral representation one can get the Hamiltonian corresponding to the asymmetrical Dirac spectra: $\widehat H=\hbar v_+k\left|\mathbf k+ \right\rangle-\hbar v_-k\left|\mathbf k- \right\rangle$. Substituting Eq.~(\ref{k}) we obtain $\widehat H=\hbar k(v_+ - v_-)/2 +\hbar\boldsymbol\sigma\mathbf k(v_+ + v_-)/2$. Then the velocity operator will be $\widehat v_x=(v_+ - v_-)/2 + \sigma_x(v_+ + v_-)/2$, and the integral in Eq.~(\ref{s_inter_B}) has the form $\int_{\Omega_{3\rm D}}\frac{\left|\left\langle \mathbf k+\right|\widehat v_x\left|\mathbf k-\right\rangle\right|^2}{4\pi}=\frac{2}{3}\left(\frac{v_+ + v_-}{2}\right)^2$. Denoting in Eq.~(\ref{s_inter_B}) $v_-\equiv v_F$, $\gamma\equiv(v_+/v_-+1)\big/2$ and $E_-\equiv-E$, then $E_+ =Ev_+/v_-$, and we get: 
\begin{equation}
\sigma_{\rm inter}(\omega)=\frac{-ie^2g\omega}{3\pi^2\hbar v_F}\gamma \int_{0}^{\infty}\left(\frac{n(Ev_+/v_-)-n(-E)}{\hbar^2(\omega+i0)^2-4E^2\gamma^2}\right)EdE.
\end{equation}
Resolving the singularity $E=\hbar\omega/2\gamma$ in the same way as we have done in Appendix \ref{Appendix_A} we obtain: 
\begin{equation}\textbf{}\sigma_{\rm inter}(\omega)=\frac{ie^2g\omega}{3\pi^2\hbar v_F}\gamma\left[-\frac{\pi i}{2}\frac{\widetilde G(\hbar\omega/2\gamma)}{4}+\int_{0}^{\infty}\left(\frac{\widetilde G(E)-\widetilde G(\hbar\omega/2\gamma)}{\hbar^2\omega^2-4E^2\gamma^2}\right) EdE\right],
\end{equation}
where $\widetilde G(E)=n(-E)-n(Ev_+/v_-)$. Thus the real and imaginary parts of the longitudinal dynamic conductivity in the case of the asymmetrical Dirac spectra are written as [notations are the same as for Eqs.~(\ref{ResigT_A}) and (\ref{ImsigT_A})]: 
\begin{align}
\textup{Re}\, \sigma\! \left(\Omega\right )=&\,\frac{e^2}{\hbar}\frac{gk_F}{24\pi}\Omega\gamma\widetilde G\left(\Omega/2\gamma\right),
\\ 
\textup{Im}\, \sigma\! \left(\Omega\right)=&\,\frac{e^2}{\hbar}\frac{gk_F}{24\pi^2}\biggl[\frac{4}{\Omega}\left( 1+\frac{\pi^2}{3}\left(\frac{T}{E_F}\right)^{2}\right)+ 8\Omega\gamma\int_{0}^{\varepsilon _c}\left (\frac{\widetilde G\left ( \varepsilon \right )-\widetilde G\left (\Omega/2\gamma\right )}{\Omega^2-4\varepsilon ^2\gamma^2}\right )\varepsilon d\varepsilon\biggr], 
\end{align}	
At the low-temperature limit  $\widetilde G\left(\Omega/2\gamma\right)\to \rm{\theta}(\Omega-2\gamma)$, and we obtain the similar expressions as for the symmetrical case [Eqs.~(\ref{Resig_A}) and (\ref{Imsig_A})], but with the $\gamma$ factor:
\begin{align}
\textup{Re}\, \sigma\! \left(\Omega\right)=&\,\frac{e^2}{\hbar }\frac{gk_F}{24\pi}\Omega\gamma\,\rm{\theta}(\Omega-2\gamma), \label{Resig_B}
\\
\textup{Im}\, \sigma\! \left(\Omega\right)=&\,\frac{e^2}{\hbar}\frac{gk_F}{24\pi^2}\left[\frac{4}{\Omega }-\frac{\Omega}{\gamma}\ln\left(\frac{4\varepsilon_c^2\gamma^2}{\left|\Omega^2-4\gamma^2\right|}\right)\right]. 
\end{align}

\begin{thebibliography}{85}
\bibitem{Ec}
According to the data from Ref.~\cite{Borisenko} [see Fig. 2(a) therein] we take the Fermi level  $E_F =0.15$eV and the cutoff energy $E_c=0.6$eV -- $E_F=0.45$eV, giving $\varepsilon_c=E_c/E_F=3$.
\end{thebibliography}
\end{document}